# CONTROL SYSTEM OF THE BEPCII

J. Zhao, C.H. Wang, X.C. Kong, G. Lei, S.F. Xu, Q. Le
IHEP Beijing, 100039 P.R. China

Abstract

Recently the Chinese Academy of Sciences has chosen BEPCII as the future development of the BEPC, i.e. upgrade of both the machine and detector. The luminosity of the machine is expected to increase to $1.0 \times 10^{33} cm^{-2} s^{-1}$. The project will be started at the beginning of 2002 and finished with in 3-4 years. The BEPC control system was built in 1987 and upgraded in 1994. According to the design of the BEPCII, a double ring schema will be adopted and a number of new devices will be added in the system. The existing control system has to be upgraded. The BEPCII will be distributed architecture and developed by EPICS. We are going to apply the standard hardware interfaces and mature technologies in the system. A number of VME IOCs will be added in the system and the feildbus, PLCs will be used as device control for some kind of equipment. We will keep the existing system in use, such as CAMAC modules and PC front-ends, and merge it into EPICS system. Recently the development of the prototype is in progress. This paper will describe the system design and development issues.

## 1 PROJECT OF BEPCII

The Beijing Electron Positron Collider (BEPC) was constructed for both high energy physics and synchrotron radiation researches. BEPC accelerators consist of a 202 m long electron-positron linac injector; a storage ring with a circumference of 240.4 m, and in connection with each other, 210 m transport lines. There are two interaction points on the storage ring, and a detector, the Beijing Spectrometer. The Beijing Synchrotron Radiation Facility has 9 beamlines and 12 experimental stations. As an unique e+,e- collider operating in the τ-charm region and the first SR source in China, the machine has been well operated for over 10 years since it was put into operation in 1989.

To upgrade the BEPC, known as BEPCII, i.e. increasing its luminosity to $1.0 \times 10^{33} cm^{-2} s^{-1}$ with the major upgrades in both machine and the detector. Running at the resonance peaks of J/ψ and ψ', the BEPCII could provide data samples of J/ψ and ψ´ with good statistics for many important physics.[1] To reach the goal of the higher luminosity, the double ring schema will be adopted and many new equipment will be installed in the BEPCII, such as superconducting RF cavities, superconducting magnets, new BPMs and beam feedback system. The design goal of the BEPCII is in table1.

Table1 The design goal of the BEPCII

| Beam Energy range | 1- 2.8GeV |
|---|---|
| Beam energy | 1.55 GeV |
| Beam current | 1.116A |
| Luminosity | $1.0 \times 10^{33}\ cm^{-2} s^{-1}$ @1.55GeV |
| Injection from linac | Full energy injection |
| Dedicated SR operation | 250ma @ 2.5GeV 150mA @ 2.8GeV |

## 2 CURRENT SYSTEM

The current control system of the BEPC was built in 1987, which was transferred from SLAC New Spear system and upgraded in 1994. A VAX4500 computer with the CAMAC hardware controls most of the equipment on the storage ring, such as the magnet power supplies, the RF cavities and the vacuum equipment. There are about 300 CAMAC modules and 4,000 channels in the system. Several sub-control systems are PC based system including the beam diagnostic system, the injection power supply and the linac system. Two VAX workstations serve as the console and all of the machines are connected with 100Mbps Ethernet. On the software side, the system is an in-house made database driven system. And the applications for accelerator operation with its OPI are closely linked to the real-time database. However, the accelerator commissioning model of the BEPCII is different from BEPC, so the accelerator commissioning programs have to be transferred and modified from other laboratories. In this case, the existing real-time database does not match the new transferred applications and the old OPI written by FORTH language will not be used in future. The control system has to be upgraded.

When the system upgrade, we will utilize the existing equipment of the system, such as the CAMAC modules and PC based subsystems. The standard hardware interfaces should be applied in the system so that it could be an open and standard system. With regard to the software development environment, the EPICS will be a good choice for it is a mature software package and widely used in the world. The cost-performance of the system should be considered for the limited budget.

## 3 SYSTEM DEVELOPMENT

### 3.1 Software Engineering

Project management is very important for the development of a control system. The BEPCII control system has to be delivered on time and within the budget and it should meet the requirement of accelerator physicists. The standard of software engineering will be used in the project, which is based on software life cycle reference model. It includes several phases: the user requirement phase, system analysis and software requirement analysis, system design, coding and testing, installation and put into operation. The documentation and review for the major phases will be done. A software tool "Project 2000" will be used to manage the project BEPCII. And we are going to produce Chinese template of the documents for the project management that will be used by our users and developers.

Now the user requirement phase is in progress. A detailed system design will be delivered in next March, and a prototype of the control system will be built at same time. From beginning of 2003 we will spend two years for system construction, and one year for system installation. We hope the BEPCII will finish its' commissioning in December of 2006.

### 3.2 Development Tool

To develop a large scale control system, a SCADA (Supervisory Control and Data Acquisition) tool kit should be used. Currently there are many commercial SCADA products in the world market and most of them are running on the PC and Windows NT platform, which mainly support PLC hardware. The EPICS can be considered a non-commercial SCADA tool, which was first developed by LANL and ANL and it is widely used in the accelerator area. We are going to integrate the BEPCII control system by the EPICS after evaluating the commercial SCADA products and EPICS [2]. The benefit of using EPICS is that a lot of applications for accelerator commissioning could be shared, it strongly supports the VME hardware and it is easy to get technical support from many HEP institutes in the world, such as KEK-B, SLAC, APS, LANL and BESSYII. Some independent and slow control system could be developed by a commercial SCADA product.

## 4 SYSTEM ARCHITECTURE

The BEPCII control system will adopt a distributed architecture, called "standard model". Logically, the system is structured with three levels, which are presentation layer, process control layer and device interface layer [3].

As shown in Figure 1, several PCs, the cost-effective equipment, and SUN workstations will be used as operator console at the presentation layer with the EPICS/OPI, which is a friendly graphic man-machine interface. There is a server machine in the top layer installing Oracle database, which provides data logging and analysis service, the accelerator commissioning support, and general computing resources.[4]

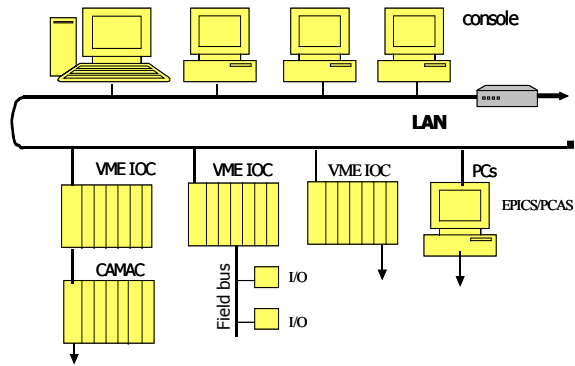

Figure1 System architecture

In the process control layer, the PowerPC microprocessors running VxWorks operating system will be used as EPICS IOCs. In some existing subsystems the PCs are still used as the front-ends. The VAX4500 computer will be eliminated. The majority real-time tasks will run on the front-ends and the raw data are stored in the IOC real-time database. The standard 100Mbit/sec switch Ethernet with TCP/IP protocol is used as a LAN, which provides access to the distributed computers.

The device interface layer provides interface to the hardware, either as separate modules or as intelligent controllers. The fieldbuses will serve data exchange with PLCs, intelligent controllers, and remote I/O modules.

## 5 SUBSYSTEMS

There are seven subsystems in current BEPC control system: the magnet power supply system in the storage ring and the transport lines, the radio frequency system, the vacuum system, the injection system, the beam diagnostic system and the linac system.

To upgrade the control system and meet the requirement of the BEPCII, the new parts of the system will be built with EPICS and some existing parts should be merged into the EPICS system. (See figure1)

The BEPC has a single ring and the BEPCII will adopt double ring schema to obtain higher luminosity. The number of the magnet power supplies will be increased on the storage ring (see Table2) and the superconducting RF cavities, new BPMs and beam feedback system will be added in the BEPCII. These equipment will be controlled by tens EPICS IOCs which consists of VME crates, PowerPCs, DSPs and VME I/O modules.

Table2 Number of magnet power supplies on SR

| Power supply for | BEPCII | BEPC |
|---|---|---|
| Bending magnets | 6 | 1 |
| Quadruple magnets | 144 | 21 |
| Correctors | 160 | 64 |
| Superconducting Mag. | 10 | 0 |
| Sextuple magnets | 4 | 4 |

The vacuum equipment and magnet power supplies of Linac accelerator will be controlled by PLCs and remote I/O modules, which are connected to the EPICS IOCs via feildbus. The ControlNet, CANbus are the candidate of the feildbuses.[5]

There are several PCs for beam diagnostic system running applications developed by Labview. The EPICS/PCAS will be installed on these PCs to merge the existing system to EPICS system.

Some parts of the control system still use CAMAC I/O modules. We are going to add VME IOC in the local control area and connect the CAMAC system to EPICS system with VME-CAMAC interfaces.

Regards some slow controls, such as cooling water system, we are going to develop the independent system with commercial SCADA products, because it supports PLCs, various sensor and controllers with industrial standards.[6]

## 6 APPLICATIONS

There are two kinds of applications in the system. One is application of device control; the other is application for accelerator commissioning. For the device control, we have to configure our own IOC database firstly. If need, there are some I/O drivers have to be developed for the special hardware interface. The graphic man-machine interface will be developed by EPICS tool DM2K. A lot of applications for device control should be developed by EPICS tools or written with C/C++, Tcl/tk or SNL language of EPICS. The accelerator commissioning software is considered to transfer from other laboratories, such as KEK-B or PEP-II and it has to be modified based on the physical requirement of the BEPCII.

## 7 DATABASES

There are two kinds of database in the system, one is a distributed real-time database running in EPICS IOCs to store raw data; the other is a relational database Oracle, which will be installed on a server machine to store static and dynamic data including system configuration data, machine parameters, historical and alarm data etc. And both of the two databases provide user interface on the Web page and publish the running information. The first thing for us is to make signal naming convention used in the databases and create the database structures.

## 8 TIMING SYSTEM

The timing system is to synchronize all of the relevant components in the accelerator of the BEPCII complex. Since the RF frequency of the BEPCII will be changed from 200MHZ to about 500MHZ, so the current timing system has to be upgraded. Except the hardware based fast timing system, the software based timing system might be considered in some synchronization area.

## 9 CONCLUSION

The project BEPCII will be started next year. In order to build the control system with EPICS and develop basic applications, a prototype system should be created firstly. After making a great effort, we think the upgraded system can meet the physical requirement of the BEPCII and it will be an advanced, flexible and reliable system.

## 10 ACKNOWLEDGEMENT

The authors would like to thank all of the people and friends who have given us a lot of helps and advices in the system design stage.

## REFERENCES

[1] "Feasibility Study Report on BEPCII ", Edition1, IHEP April 2001
[2] "EPICS Release R3.14", APS Argonne.
[3] Tadahiko Katoh, "Design and Construction Accelerator Control System" KEKB control group, 28 Aug. 2000.
[4] J.Zhao et al, "Preliminary Design of the Control System for SSRF and BTCF" ICALEPCS'97 p56 – p58, Nov. 1997 Beijing, China.
[5] http://www.sns.bnl.gov/epics/cnet/
[6] A. Daneels , W. Salte "What is SCADA" ICALEPCS'99 p339-343 Oct. 1999 Trieste, Italy.